\documentclass{llncs}
\usepackage{amsmath, amsfonts, graphicx, booktabs, tabularx, multirow, siunitx, xcolor, soul}
\usepackage{hyperref}

\graphicspath{{./figures/}}
\begin{document}

\title{SegNAS3D: Network Architecture Search with Derivative-Free Global Optimization\\ for 3D Image Segmentation\thanks{This paper was accepted by the International Conference on Medical Image Computing and Computer-Assisted Intervention -- MICCAI 2019.}}
\author{Ken C. L. Wong, Mehdi Moradi}
% index{Wong, Ken C. L.}
% index{Moradi, Mehdi}
\institute{IBM Research -- Almaden Research Center, San Jose, CA, USA\\
\email{\{clwong, mmoradi\}@us.ibm.com}
}

\maketitle              % typeset the title of the contribution

\begin{abstract}
Deep learning has largely reduced the need for manual feature selection in image segmentation. Nevertheless, network architecture optimization and hyperparameter tuning are mostly manual and time consuming. Although there are increasing research efforts on network architecture search in computer vision, most works concentrate on image classification but not segmentation, and there are very limited efforts on medical image segmentation especially in 3D. To remedy this, here we propose a framework, SegNAS3D, for network architecture search of 3D image segmentation. In this framework, a network architecture comprises interconnected building blocks that consist of operations such as convolution and skip connection. By representing the block structure as a learnable directed acyclic graph, hyperparameters such as the number of feature channels and the option of using deep supervision can be learned together through derivative-free global optimization. Experiments on 43 3D brain magnetic resonance images with 19 structures achieved an average Dice coefficient of 82\%. Each architecture search required less than three days on three GPUs and produced architectures that were much smaller than the state-of-the-art manually created architectures.
\end{abstract}
\section{Introduction}

Although deep learning has largely reduced the need for manual feature selection in image segmentation \cite{Conference:Cicek:MICCAI2016,Conference:Milletari:3DV2016}, days to weeks are still required to manually search for the appropriate architecture and hyperparameters. To further reduce human workloads, network architecture search (NAS) has been proposed for image classification in the computer vision community to automatically generate network architectures. In \cite{Journal:Zoph:arXiv2016}, a recurrent network was used to generate the model descriptions of neural networks, which was trained with reinforcement learning on 800 GPUs to learn architectures from scratch. In \cite{Conference:Zhong:CVPR2018}, a block-wise network generation pipeline was introduced to automatically build networks using the Q-learning paradigm with tremendous increase of search efficiency.

Although these works are promising, efforts on NAS for medical image segmentation are very limited especially in 3D. In \cite{Conference:Mortazi:MLMI2018}, the policy gradient reinforcement learning has been used to learn the kernel size and the number of feature channels of each convolutional layer of a custom network architecture for 2D medical image segmentation. Without learnable layer interconnections, this framework mainly performs hyperparameter tuning rather than architecture search, and the computational complexity is infeasible for 3D image segmentation. In fact, the computational requirements for 3D images are much higher than 2D images. Furthermore, multiple local optima can be expected in the architecture search space but they are not handled by most frameworks. Therefore, developing an efficient NAS framework for 3D images is a very challenging task.

In view of these issues, here we propose a NAS framework, SegNAS3D, for 3D image segmentation with two key contributions. \textbf{I)} For computational feasibility, inspired by \cite{Journal:Liu:arXiv2018}, the overall network architecture is composed of repetitive block structures, with each block structure represented as a learnable directed acyclic graph. Different from \cite{Journal:Liu:arXiv2018}, the interconnections among block structures are also modeled as learnable hyperparameters for a more complete network architecture search. \textbf{II)} By constructing the hyperparameter search space with continuous relaxation and handling untrainable situations such as the out-of-memory (OOM) error, derivative-free global optimization is applied to search for the optimal architecture. To the best of our knowledge, this is the first work of network architecture search for 3D image segmentation with global optimization. Experiments on 43 3D brain magnetic resonance (MR) images with 19 anatomical structures achieved an average Dice coefficient of 82\%. Each architecture search required less than three days on three GPUs, and the resulted networks were much smaller than the V-Net \cite{Conference:Milletari:3DV2016} on the tested dataset.

% ------------------ figure -------------------------
\begin{figure}[t]
    \centering
    \begin{minipage}[b]{1\linewidth}
      \centering
      \includegraphics[width=1\linewidth]{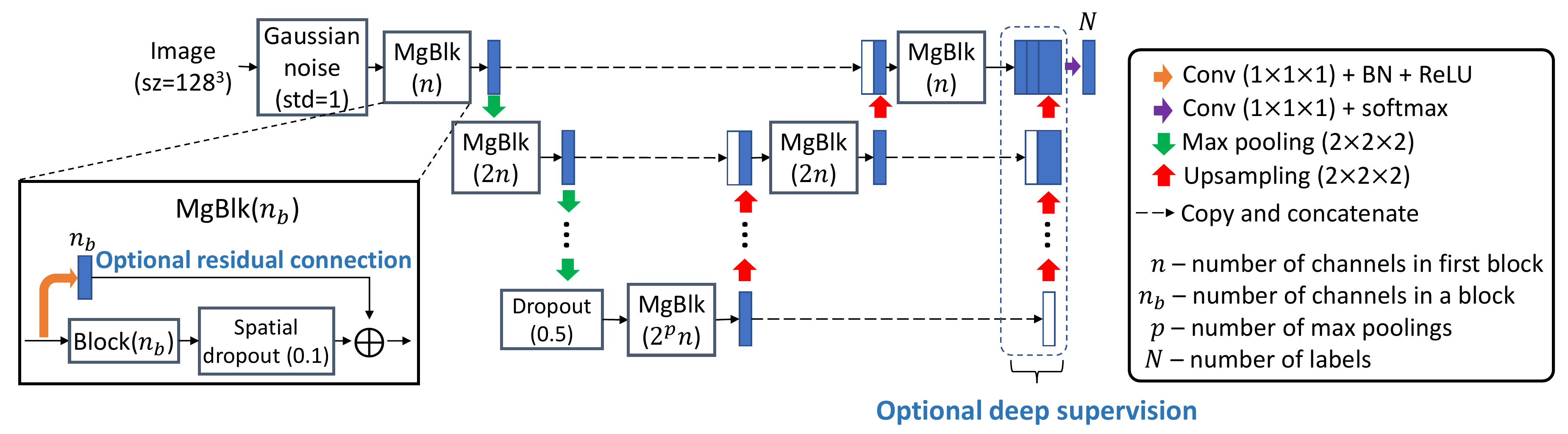}
    \end{minipage}
    \caption{Overall architecture. Blue and white boxes represent operation outputs and copied data, respectively. A MegaBlock (MgBlk) comprises a learnable block in Section \ref{sec:block} with spatial dropout and an optional residual connection. $n$, $p$, and the options of using deep supervision and block residual connections are learnable hyperparameters.}
    \label{fig:architecture}
\end{figure}
% ---------------------------------------

\section{Methodology}

For computational feasibility, inspired by \cite{Journal:Liu:arXiv2018,Conference:Zhong:CVPR2018}, the segmentation network architecture comprises two key components: the building blocks and their interconnections (Fig. \ref{fig:architecture}). A building block comprises various deep-learning layers such as convolution and batch normalization, whose pattern is repeatedly used in the overall network. The residual units of the ResNet \cite{Conference:He:ECCV2016} are good examples. The building blocks are connected together to form the network architecture. For classification networks, the blocks are usually cascaded with pooling layers in between \cite{Conference:He:ECCV2016,Conference:Szegedy:AAAI2017}. For segmentation networks, there are more variations of how different blocks are connected \cite{Journal:Yu:arXiv2015,Conference:Cicek:MICCAI2016,Conference:Milletari:3DV2016}.

% ------------------ figure -------------------------
\begin{figure}[t]
    \centering
    \begin{minipage}[b]{1\linewidth}
      \centering
      \includegraphics[width=1\linewidth]{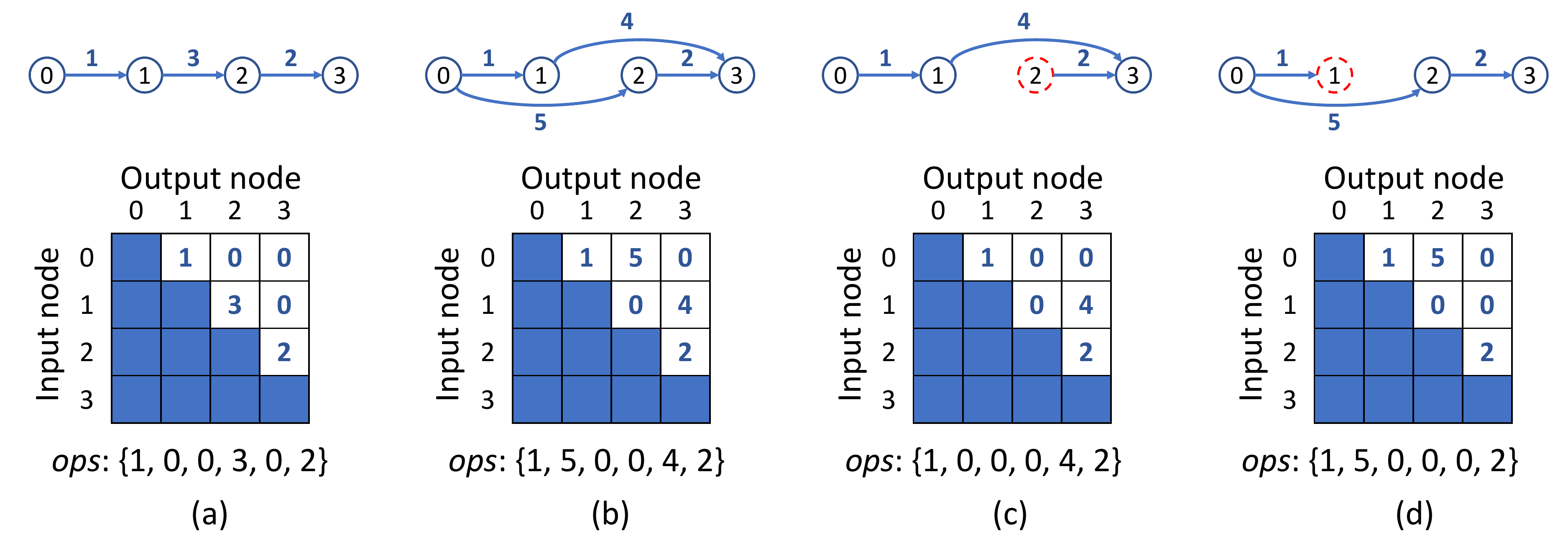}
    \end{minipage}
    \caption{Examples of upper triangular operation matrices and the corresponding block structures for a directed acyclic graph with four nodes. Each integer matrix element represents an operation in Table \ref{table:operations} and $ops$ represents the corresponding set of operations. (a) The simplest block structure. (b) A more complicated block structure with multiple nodal inputs and outputs. (c) and (d) Illegal block structures with node 2 as a source and node 1 as a sink, respectively.}
    \label{fig:block_examples}
\end{figure}
% ---------------------------------------

% ------------------ Table -------------------------
\begin{table}[t]
\caption{Block operations and their corresponding numbers. Conv($k$, $d$) represents a $k \times k \times k$ convolution with dilation rate $d$. $d = 1$ means no dilation. Each convolution is followed by batch normalization and ReLU activation.}
\label{table:operations}
\fontsize{6}{7}\selectfont
\centering
\begin{tabular*}{\textwidth}{@{\extracolsep{\fill}}lllllll}
\toprule
0 & 1 & 2 & 3 & 4 & 5 & 6 \\
\midrule
None & Conv(1, 1) & Conv(3, 1) & Conv(5, 1) & Conv(3, 2) & Conv(5, 2) & Skip connection \\
\bottomrule
\end{tabular*}
\end{table}
% --------------------------------------------------

\subsection{Block Structure}
\label{sec:block}

Inspired by \cite{Journal:Liu:arXiv2018}, a block is represented as a directed acyclic graph. Each node represents a feature map (tensor) and each directed edge represents an operation (e.g. convolution). Here we represent the graph as an upper triangular operation matrix which contains all operations among nodes (Fig. \ref{fig:block_examples}). The rows and columns of the matrix represent the input and output nodes, respectively, with nonzero elements represent operation numbers (Table \ref{table:operations}). There are two types of nodes crucial for building trainable networks. \textbf{1)} Source: a node that does not have parents in a block. \textbf{2)} Sink: a node that does not have children in a block. In a block, only the first node can be the source and the last node can be the sink as they are connected to other blocks. A network cannot be built if there are sources or sinks as the intermediate nodes. Therefore, the simplest block structure can be represented by a ``shifted'' diagonal matrix (Fig. \ref{fig:block_examples}(a)), and more complicated structures can also be achieved (Fig. \ref{fig:block_examples}(b)). With the matrix representation, a source and a sink can be easily identified as the column and the row with all zeros, respectively (Fig. \ref{fig:block_examples}(c) and (d)).

The block operations and the corresponding numbers are shown in Table \ref{table:operations}. The operations include convolutions with different kernel sizes ($k = 1, 3, 5$) and dilation rates ($d = 1, 2$) for multi-scale features \cite{Journal:Yu:arXiv2015}. Each convolution is followed by batch normalization and ReLU activation. Skip connection which allows better convergence is also included. Outputs from different nodes are combined by summation as concatenation mostly led to the OOM error in our experiments. The number of nodes ($nodes$) in a block is also a learnable hyperparameter. To reduce the complexity of architecture search, all blocks in a network share the same operation matrix, with the numbers of feature channels systematically assigned based on the number of feature channels of the first block (Section \ref{sec:architecture}).

\subsection{Network Architecture and Block-Connecting Hyperparameters}
\label{sec:architecture}

Although there are multiple ways to connect the blocks together for image segmentation, we adopted an architecture similar to the U-Net \cite{Conference:Cicek:MICCAI2016} and V-Net \cite{Conference:Milletari:3DV2016} as they were proposed for 3D medical image segmentation (Fig. \ref{fig:architecture}). The architecture contains the encoding and decoding paths with MegaBlocks. Each MegaBlock comprises a block in Section \ref{sec:block} with spatial dropout and an optional residual connection to reduce overfitting and enhance convergence. The number of channels is doubled after each max pooling and is halved after each upsampling. Deep supervision which allows more direct backpropagation to the hidden layers for faster convergence and better accuracy is also an option \cite{Conference:Lee:AISTATS2015}. The number of feature channels of the first block ($n$), the number of max poolings ($p$), and the options of using deep supervision ($sup$) and block residual connections ($res$) are learnable block-connecting hyperparameters.

\subsection{Global Optimization with Continuous Relaxation}
\label{sec:global_opt}

As the number of hyperparameter combinations can be huge ($>$141 millions in some of our experiments) and each corresponds to a network training, brute force search is prohibitive and nonlinear optimization is required. Compared with discrete optimization, there are many more continuous optimization algorithms available especially for derivative-free global optimization \cite{Book:Conn:DFO2009}. Therefore, similar to \cite{Journal:Liu:arXiv2018}, continuous relaxation is used to remove the integrality constraint of each parameter. This also allows us to introduce non-integral hyperparameters such as the learning rate if desired. Different from \cite{Journal:Liu:arXiv2018} which formulated the problem for local gradient-based optimization, we use derivative-free global optimization. This is because it is nonoptimal to compute gradients of the discontinuous objective function, and multiple local minima can be expected. Handling of untrainable situations is also simpler without gradients.

Let $\mathbf{x} \in \mathbb{R}^{n_h}$ be a vector of $n_h$ hyperparameters after continuous relaxation. We use $\lfloor \mathbf{x} \rfloor$ (floor of $\mathbf{x}$) to construct a network architecture. Therefore, the objective function is a discontinuous function in a bounded continuous search space which can be better handled by derivative-free global optimization. The objective function $f = - \ln (Dice)$ is used, where $Dice$ is the validation Dice coefficient. The derivative-free global optimization algorithm ``controlled random search'' (CRS) \cite{Journal:Kaelo:JOTA2006} is used as it provides effective search with good performance among tested algorithms. CRS starts with a population of sample points ($10\times(n_h+1)$) which are gradually evolved by an algorithm that resembles a randomized Nelder-Mead algorithm. Each search stops after 300 iterations.

Several issues need to be handled for effective and efficient search. Firstly, to handle hyperparameters of illegal block structures (Section \ref{sec:block}) and OOM errors, we assign them an objective function value $\lceil \max f \rceil$, which is 10 by clipping the minimum value of $Dice$ as $10^{-4}$. This tells the optimization algorithm that these situations are worse than having the worst segmentation. Secondly, as multiple $\mathbf{x}$ contribute to the same $\lfloor \mathbf{x} \rfloor$, we save each $\lfloor \mathbf{x} \rfloor$ and the corresponding $f$ to avoid unnecessary training for better efficiency.

\subsection{Training Strategy}

In each network training, image augmentation with rotation (axial, $\pm$\ang{30}), shifting ($\pm$20\%), and scaling ([0.8, 1.2]) is used, and each image has an 80\% chance to be transformed. The optimizer Nadam is used for fast convergence with the learning rate as 10$^{-3}$. The exponential logarithmic loss with Dice loss and cross-entropy is used \cite{Conference:Wong:MICCAI2018}. The IBM Power System AC922 equipped with NVLink for enhanced host to GPU communication was used. This machine features NVIDIA Tesla V100 GPUs with 16 GB memory, and three of these GPUs were used for multi-GPU training with a batch size of three and 100 epochs.

% ------------------ Table -------------------------
\begin{table}[t]
\caption{Variations of the proposed framework with different learnable hyperparameters and their lower and upper bounds. The effective set of integers of each half-open interval [$a$, $b$) is \{$a$, $\ldots$, $b-1$\}. For bounds [0, 2), \{0, 1\} means \{Disable, Enable\}. The upper bound of $nodes$ determines the number of block-operation hyperparameters ($ops$) required. For example, $nodes$ with bounds [2, 5) requires six $ops$ to fill a 4$\times$4 upper triangular matrix. Scalars in red are fixed. For SegNAS$_4$, $ops$ of \{2, 0, 2\} represents two cascaded Conv(3, 1) in Table \ref{table:operations}.}
\label{table:hyperbounds}
\fontsize{6}{7}\selectfont
\centering
\begin{tabular*}{\textwidth}{@{\extracolsep{\fill}}lllllll}
\toprule
& \multicolumn{4}{c}{Block-connecting hyperparameters} & \multicolumn{2}{c}{Block structures} \\
\cmidrule{2-5}\cmidrule{6-7}
& $n$ & $p$ & $sup$ & $res$ & $nodes$ & $ops$  \\
\midrule\relax
SegNAS$_{11}$ & [8, 33) & [2, 6) & [0, 2) & [0, 2) & [2, 5) & [0, 7) ($6\times$) \\
SegNAS$_{4}$ & [8, 33) & [2, 6) & [0, 2) & [0, 2) & {\color{red}\textbf{3}} & {\color{red}\textbf{\{2, 0, 2\}}} \\
SegNAS$_{7}$ & {\color{red}\textbf{16}} & {\color{red}\textbf{4}} & {\color{red}\textbf{0}} & {\color{red}\textbf{1}} & [2, 5) & [0, 7) ($6\times$) \\
\bottomrule
\end{tabular*}
\end{table}
% --------------------------------------------------

\section{Experiments}

\subsection{Data and Experimental Setups}

We validated our framework on 3D brain MR image segmentation. A dataset of 43 T1-weighted MP-RAGE images from different patients was neuroanatomically labeled to provide the training, validation, and testing samples. They were manually segmented by highly trained experts, and each had 19 semantic labels of brain structures. Each image was resampled to isotropic spacing using the minimum spacing, zero padded, and resized to 128$\times$128$\times$128.

Three sets of dataset splits were generated by shuffling and splitting the dataset, with 50\% for training, 20\% for validation, and 30\% for testing in each set. The training and validation data were used during architecture search to provide the training data and the validation Dice coefficients for the objective function. The testing data were only used to test the optimal networks after search. Three variations of our proposed framework were tested (Table \ref{table:hyperbounds}). SegNAS$_{11}$ optimizes both block structures and their interconnections. SegNAS$_{4}$ optimizes only the block-connecting hyperparameters with fixed block structures. SegNAS$_{7}$ optimizes only the block structures with fixed block-connecting hyperparameters inferred from the V-Net. Note that the subscripts indicate the numbers of hyperparameters to be optimized. We performed experiments on the 3D U-Net \cite{Conference:Cicek:MICCAI2016} and V-Net \cite{Conference:Milletari:3DV2016} for comparison. The same training strategy and dataset splits were used in all experiments.

% ------------------ figure -------------------------
\begin{figure}[t]
    \centering
    \begin{minipage}[t]{0.6\linewidth}
      \centering
      \includegraphics[width=1\linewidth]{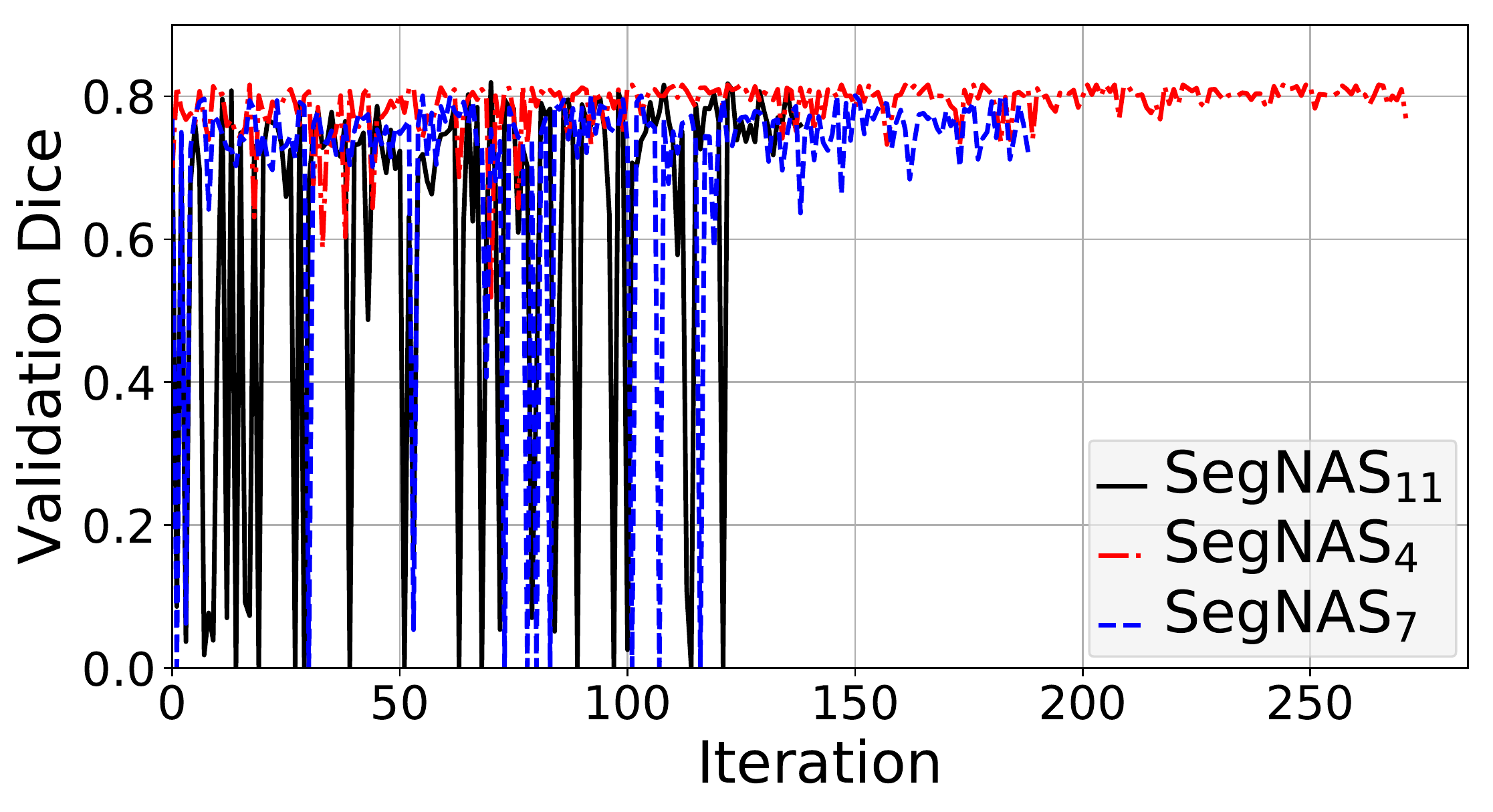}
    \end{minipage}
    \caption{Evolutions of the validation Dice coefficients during search. Examples of a dataset split. Iterations with illegal block structures and OOM errors are omitted. The effective numbers of validation Dice coefficients for SegNAS$_{11}$, SegNAS$_{4}$, and SegNAS$_{7}$ were 139, 272, and 189, respectively, out of 300 iterations.}
    \label{fig:plot_valid_dice}
\end{figure}
% ---------------------------------------

\subsection{Results and Discussion}

Examples of the evolutions of the validation Dice coefficients during search are shown in Fig. \ref{fig:plot_valid_dice}. In all tests, there were more fluctuations at the early iterations as the optimization algorithm searched for the global optimum, and the evolutions gradually converged. SegNAS$_{11}$ had the least effective number of Dice coefficients (139) as its larger number of hyperparameter combinations led to more illegal structures and OOM errors. In contrast, SegNAS$_{4}$ had the most effective number (272). We can also see that searching optimal block structures (SegNAS$_{11}$ and SegNAS$_{7}$) led to larger fluctuations, and searching only block-connecting hyperparameters (SegNAS$_{4}$) gave faster convergence.

\begin{figure}[t]
    \fontsize{6}{7}\selectfont
    \centering
    \begin{minipage}[t]{0.15\linewidth}
      \centering
      \includegraphics[width=1\linewidth]{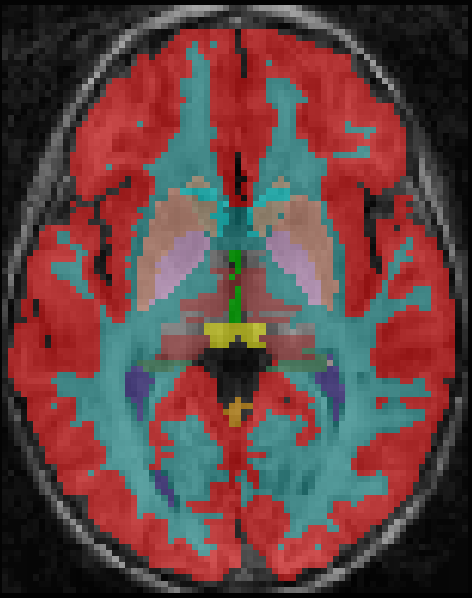} \\
      \includegraphics[width=1\linewidth]{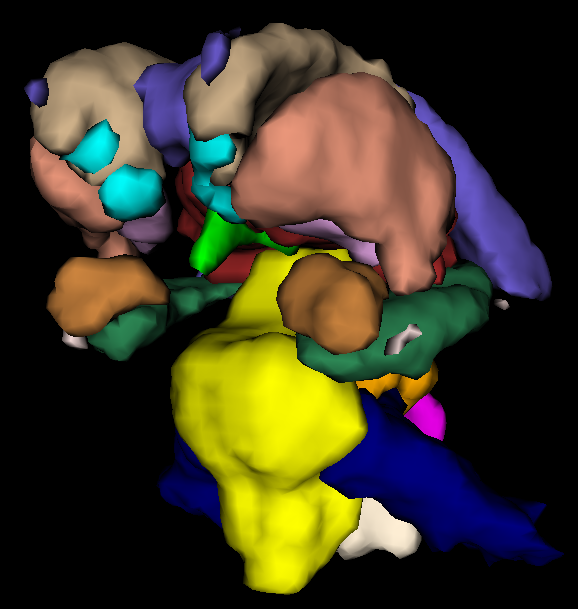} \\
      \centering{Ground truth}
    \end{minipage}
    \vrule\
    \begin{minipage}[t]{0.15\linewidth}
      \centering
      \includegraphics[width=1\linewidth]{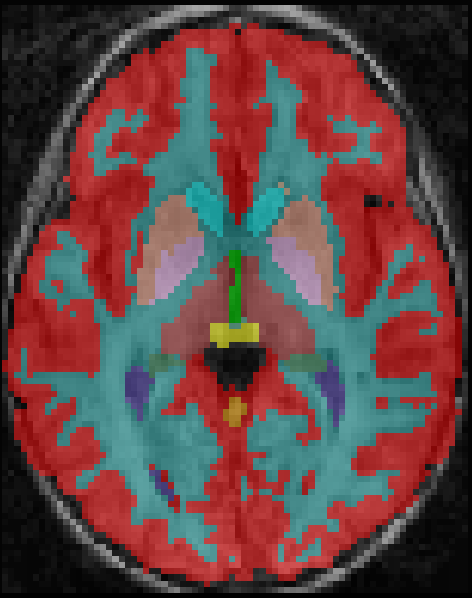} \\
      \includegraphics[width=1\linewidth]{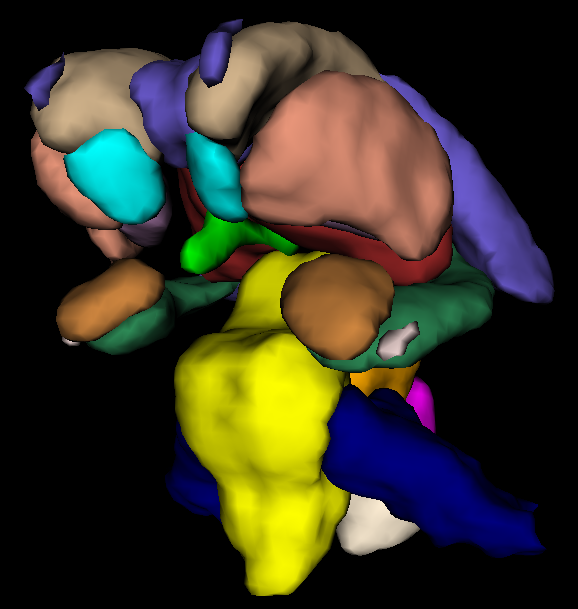} \\
      \centering{SegNAS$_{11}$ \\ Dice = 83\%}
    \end{minipage}
    \begin{minipage}[t]{0.15\linewidth}
      \centering
      \includegraphics[width=1\linewidth]{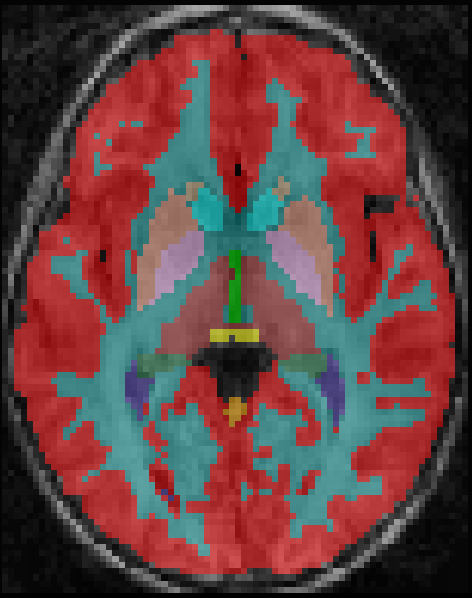} \\
      \includegraphics[width=1\linewidth]{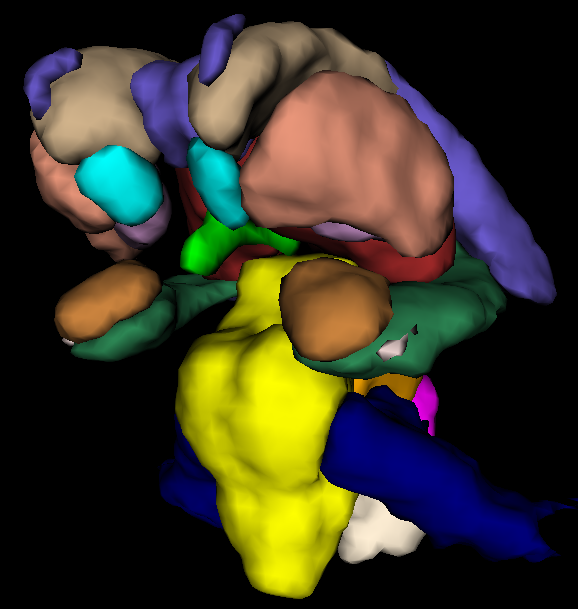} \\
      \centering{SegNAS$_{4}$ \\ Dice = 82\%}
    \end{minipage}
    \begin{minipage}[t]{0.15\linewidth}
      \centering
      \includegraphics[width=1\linewidth]{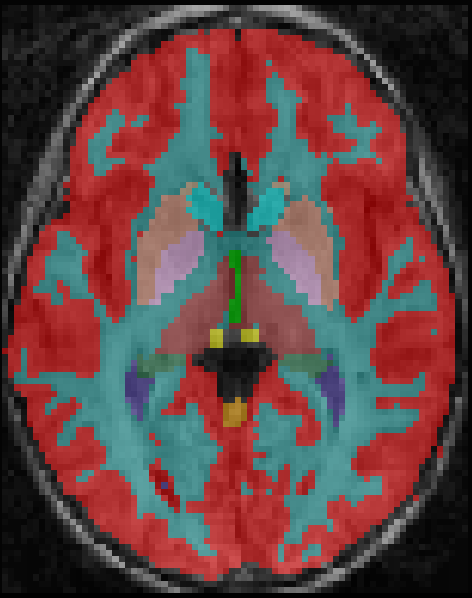} \\
      \includegraphics[width=1\linewidth]{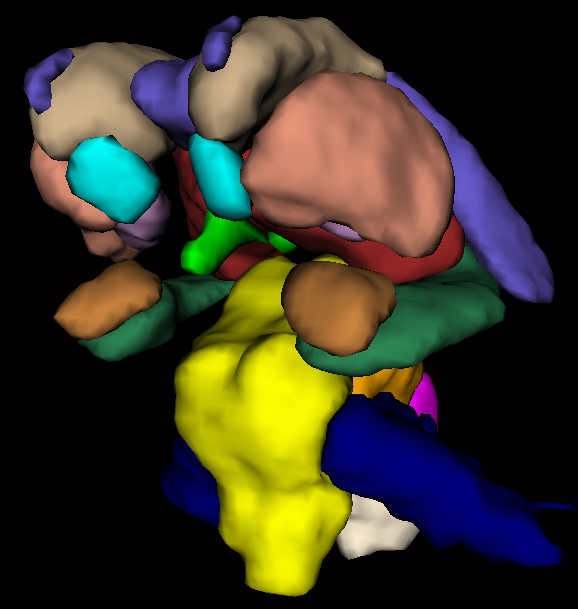} \\
      \centering{SegNAS$_{7}$ \\ Dice = 78\%}
    \end{minipage}
    \vrule\
%    \begin{minipage}[t]{0.15\linewidth}
%      \centering
%      \includegraphics[width=1\linewidth]{A_unet} \\
%      \includegraphics[width=1\linewidth]{3D_unet} \\
%      \centering{U-Net$_{1/2}$ \\ Dice = 80\%}
%    \end{minipage}
    \begin{minipage}[t]{0.15\linewidth}
      \centering
      \includegraphics[width=1\linewidth]{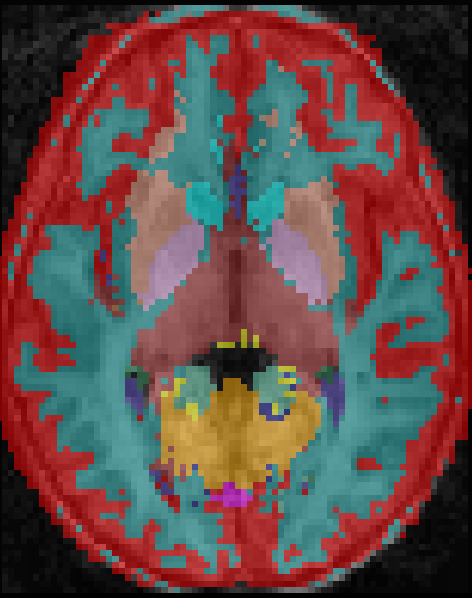} \\
      \includegraphics[width=1\linewidth]{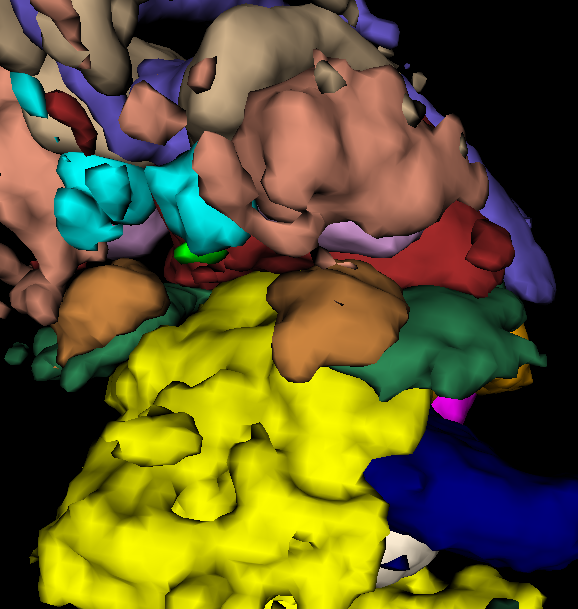} \\
      \centering{V-Net \\ Dice = 51\%}
    \end{minipage}
    \caption{Visualization of an example. Top: axial view. Bottom: 3D view with the cerebral grey, cerebral white, and cerebellar grey matters hidden for better illustration.}
    \label{fig:visualization}
\end{figure}

Table \ref{table:results} shows the average results from all three dataset splits and the optimal hyperparameters of a dataset split. The V-Net gave the lowest testing Dice coefficients and the largest model. SegNAS$_{11}$ had the best segmentation performance while SegNAS$_{4}$ produced the smallest models with fewest GPU days for comparably good performance. Among the variations, SegNAS$_{7}$ had the lowest Dice coefficients, largest models, and most GPU days. The 3D U-Net gave the OOM error and produced a larger network than SegNAS$_{11}$ and SegNAS$_{4}$. As three GPUs were used, each search required less than three days to complete. Fig. \ref{fig:visualization} shows the results of an example which are consistent with Table \ref{table:results}.

% ------------------ Table -------------------------
\begin{table}[t]
\caption{Average results of all dataset splits and the optimal hyperparameters of a dataset split (same split as Fig. \ref{fig:plot_valid_dice}). The best results are in blue and the fixed hyperparameters are in red. The testing Dice coefficients are shown. GPU days are the number of searching days multiplied by the number of GPUs (three) used. Strikethrough $ops$ of SegNAS$_{11}$ were not used to form the network because of the number of nodes (three). Please refer to Section \ref{sec:block} and \ref{sec:architecture} for the definitions of hyperparameters.}
\label{table:results}
\fontsize{6}{7}\selectfont
\centering
\begin{tabular*}{\textwidth}{@{\extracolsep{\fill}}llllllllll}
\toprule
& \multicolumn{3}{c}{Average results (mean$\pm$std)} & \multicolumn{6}{c}{Optimal hyperparameters of a search} \\
\cmidrule{2-4}\cmidrule{5-10}
& Dice (\%) & Parameters (millions) & GPU days & $n$ & $p$ & $sup$ & $res$ & $nodes$ & $ops$  \\
\midrule\relax
SegNAS$_{11}$ & {\color{blue}\textbf{81.7$\pm$0.3}} & 9.7$\pm$4.1 & 6.6$\pm$0.6 & 26 & 3 & 0 & 1 & 3 & \{2, 2, 3, \st{6, 3, 3}\} \\
SegNAS$_{4}$ &  81.0$\pm$0.5 & {\color{blue}\textbf{3.2$\pm$0.6}} & {\color{blue}\textbf{3.6$\pm$0.1}} & 21 & 3 & 1 & 0 & {\color{red}\textbf{3}} & {\color{red}\textbf{\{2, 0, 2\}}} \\
SegNAS$_{7}$ & 77.7$\pm$1.0 & 30.1$\pm$5.4 & 8.2$\pm$0.4 & {\color{red}\textbf{16}} & {\color{red}\textbf{4}} & {\color{red}\textbf{0}} & {\color{red}\textbf{1}} & 4 & \{6, 2, 3, 0, 4, 3\} \\
3D U-Net & OOM & 19.1$\pm$0.0 & --- & \multicolumn{6}{c}{---} \\
%U-Net$_{1/2}$ & 79.2$\pm$0.5 & 4.8$\pm$0.0 & --- & \multicolumn{6}{c}{---} \\
V-Net & 47.9$\pm$7.4 & 71.1$\pm$0.0 & --- & \multicolumn{6}{c}{---} \\
\bottomrule
\end{tabular*}
\end{table}
% --------------------------------------------------

% ------------------ figure -------------------------
% ---------------------------------------

Therefore, the block-connecting hyperparameters $n$, $p$, $sup$, and $res$ are more effective especially with simple block structures such as that of SegNAS$_{4}$. Searching also the block structures can improve segmentation accuracy with increased searching time and probably larger models. Searching only the block structures can lead to larger models depending on the fixed $n$, $p$ values and is not as effective. The 3D U-Net gave the OOM error because of its relatively large memory footprint (e.g. tensors of 128$\times$128$\times$128 with 64 feature channels). The segmentations of the V-Net were inaccurate probably because of insufficient training data given the number of network parameters. When we increased the amount of training data from 50\% to 70\%, the testing Dice coefficients of the V-Net increased to 68.1$\pm$2.3\%. These show the advantages of our framework as the OOM error is explicitly considered and the relation between the network size and the available data is automatically handled.

\section{Conclusion}

We present a network architecture search framework for 3D image segmentation. By representing the network architecture with learnable connecting block structures and identifying the hyperparameters to be optimized, we formulate the search as a global optimization problem with continuous relaxation. With its flexibility, we studied three variations of the framework. The results show that the block-connecting hyperparameters are more effective, and optimizing also the block structures can further improve the segmentation performance.

\bibliographystyle{splncs03}
\bibliography{Ref}

\end{document}